# A high precision TDC based on a multi-phase clock[1]

QI Zhong, MENG Xiangting, LI Deyuan, YANG Lei, YAO Zeen, LI Dongcang[2]

(School of Nuclear Science and Technology, Lanzhou University, Lanzhou 730000, China)

**Abstract:** The design of a high-precision time-to-digital converter (TDC) based on a multiphase clock implemented using a single field-programmable gate array is discussed in this paper. The TDC can increase the resolution of the measurement by using time interpolation. A phase-locked loop is used to generate four multiphase clocks whose frequencies are the same and whose phases are 0°, 45°, 90°, and 135°. In addition, the duty ratios of the four clocks are 50%. By utilizing four multiphase clocks to make up the interpolation clock, one clock period can be divided into eight uniform parts. The resolution of the TDC can be improved to 1/8 of a clock period. Furthermore, we have also designed a discriminator circuit for identifying the start and stop signals. On the basis of this circuit, the TDC can still measure the time interval of two signals when the start and stop signals are uncertain. The experimental results indicate that the time resolution of the TDC can achieve the theoretical value, and the linearity is very good. The architecture consumes fewer logic cells and is more stable.

## 1 Introduction

A high-precision time-to-digital converter (TDC) is widely used in many scientific fields such as space science[1-3], high-energy physics[4-6], and particle physics experiments. Accurate time-interval measurement technology, especially technology with picosecond resolution, is particularly important[7-8]. A high-precision TDC with picosecond resolution has been reported in an application-specific integrated circuit (ASIC). However, the design of an ASIC device is very complex, has a high cost, and has a slow turnaround time, making them particularly suitable for highly specialized applications. With the development of programmable logic device techniques, more TDCs have been implemented in field-programmable gate arrays (FPGAs) because of their flexibility, low cost, and short turnaround time compared with ASIC architectures.

There exist several techniques to measure the time difference between two subsequent digital signals. The simplest TDC is a counter triggered by a clock, and the clock frequency determines the resolution[9]. There are many approaches for improving the resolution of the time measurement, such as time interpolation, time stretching, the Vernier method, and the tapped delay line. Among

1 Supported by National Natural Science Foundation of China (11075069,11375077,11027508)

2 E-mail: pelab@lzu.edu.cn

them, time interpolation is a straightforward method for obtaining a high resolution[10-11]. Recently, the coarse–fine interpolation method has been greatly developed. The architecture of this method includes coarse and fine stages. The coarse stage contains a binary counter triggered by the global clock that records the most significant bit (the higher bits) of a time interval. The fine stage performs interpolation to obtain the least significant bit (the lower bits). The fine data indicate the digitized accurate time interval between the tested signal and the corresponding leading edge of the global clock[12-13]. The interpolation can be implemented in many ways such as a delay-locked loop (DLL) delay line and tapped delay line assembled by the carry chain in an FPGA. These methods can achieve a ten-picosecond time resolution, but the cost is more resources. Because the delay line comprises many delay cells, the time of delay cell is not uniform and also affected by the working voltage and temperature.

In this paper, we introduce a high-precision TDC used to measure a muon particle. In order to measure the track of a muon particle, a MicroMegas detector was utilized in the experiment. By using a delay line to connect to the MicroMegas detector, we can obtain the time difference between the two outputs of the delay line. Therefore, the incident position of a particle can be calculated by measuring the time difference via the TDC. We propose a new time interpolation method based on a multiphase clock. The architecture was implemented on an Altera Stratix III FPGA. Stratix III devices deliver abundant phase-locked loop (PLL) resources with up to 12 PLLs per device and up to 10 outputs per PLL. Each PLL can be used to independently program every output, to create a unique and customizable clock frequency, and to adjust the clock phase precisely. In this study, we use a PLL to generate four multiphase clocks with the same frequency multiplication, and each clock's phase has the same delay compared with the previous clock. In this case, one clock period can be divided into the same eight parts, and the time resolution can be 1/8 of a clock period. The frequency of the time clock is 1 GHz after multiplication; therefore, the resolution of the TDC is approximately 125 ps. We also describe how to measure the time interval in the case of when the start and stop signals are uncertain.

## 2 Measurement principles

The simplest time TDC is a counter, and the its principle of operation is shown in Fig.1. The time interval is measured by a counter between the rising edges of the reference clock immediately following the start and stop signals. The counter number is n; therefore, the value of the time interval is

$$t_n = T_0(n-1) \tag{1}$$

where $T_0$ is the reference clock's period. There is an error compared with the actual value of ($t_1$-$t_2$).

$t_1$ is the time difference between the rising edges of the start signal and the subsequent reference-clock edge, and $t_2$ is the time difference between the rising edges of the stop signal and the subsequent reference-clock edge. The maximum error may be one cycle of the clock. In the fine stage, utilization of the interpolation clock to measure $t_1$ and $t_2$ can decrease the error. The principle of operation of the interpolation clock is shown in the Fig.2. There are four copies of the same clock shifted by 0 °(Clock1), 45 °(Clock2), 90 °(Clock3), and 135 °(Clock4). Clock1 is used as the reference clock. The interpolation clock divides Clock1 into eight uniform parts. When the rising edges of first digital signal arrive, the phase difference between the signal and Clock1 can be measured, and $t_1$ can also be determined. In the same way, we can also obtain the value of $t_2$. The time interval is

$$t_n = T_0(n-1) + t_1 - t_2 \tag{2}$$

In this way, the time resolution of TDC can be improved to one-eighth of clock.

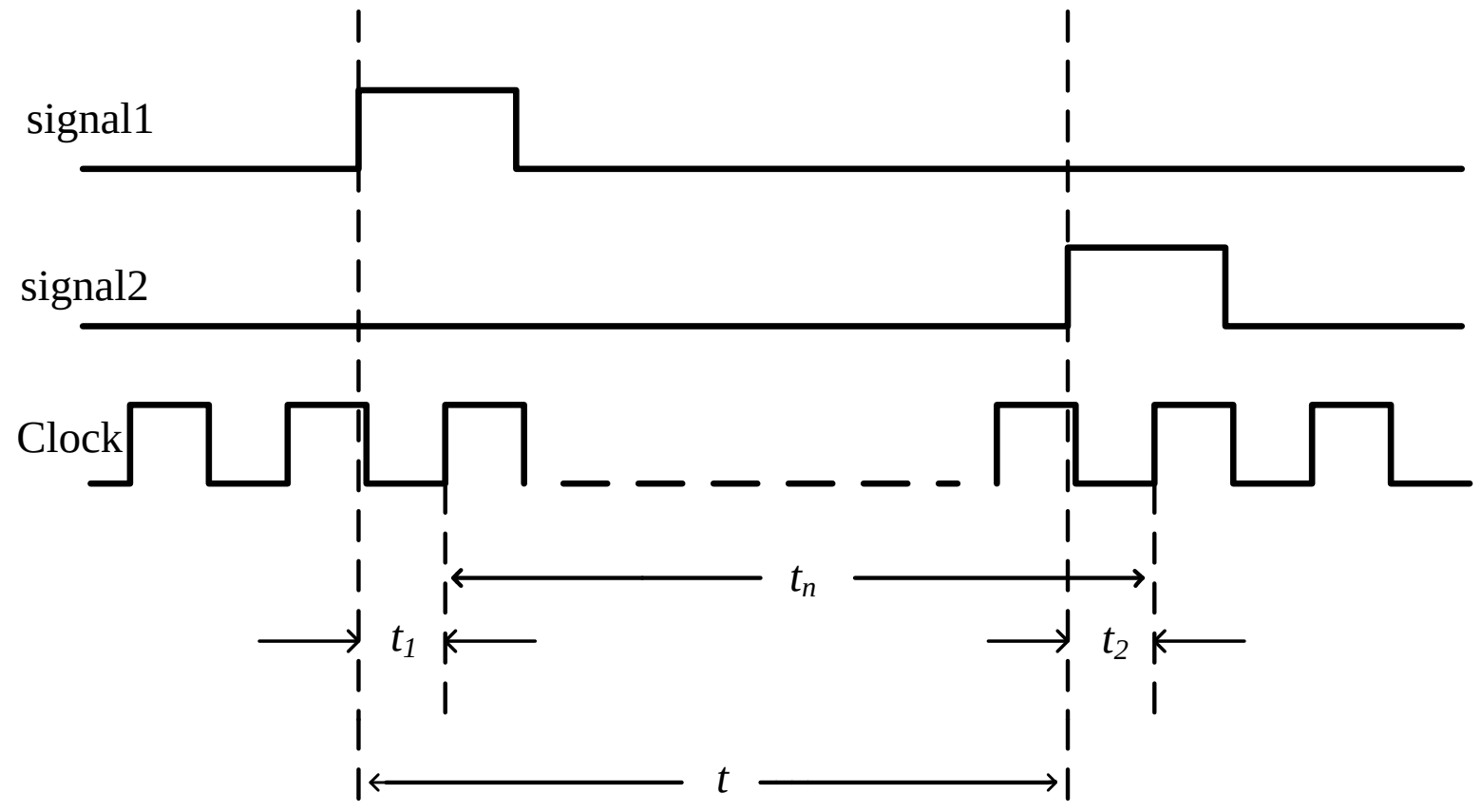

Fig.1 Principle of operation of the counter

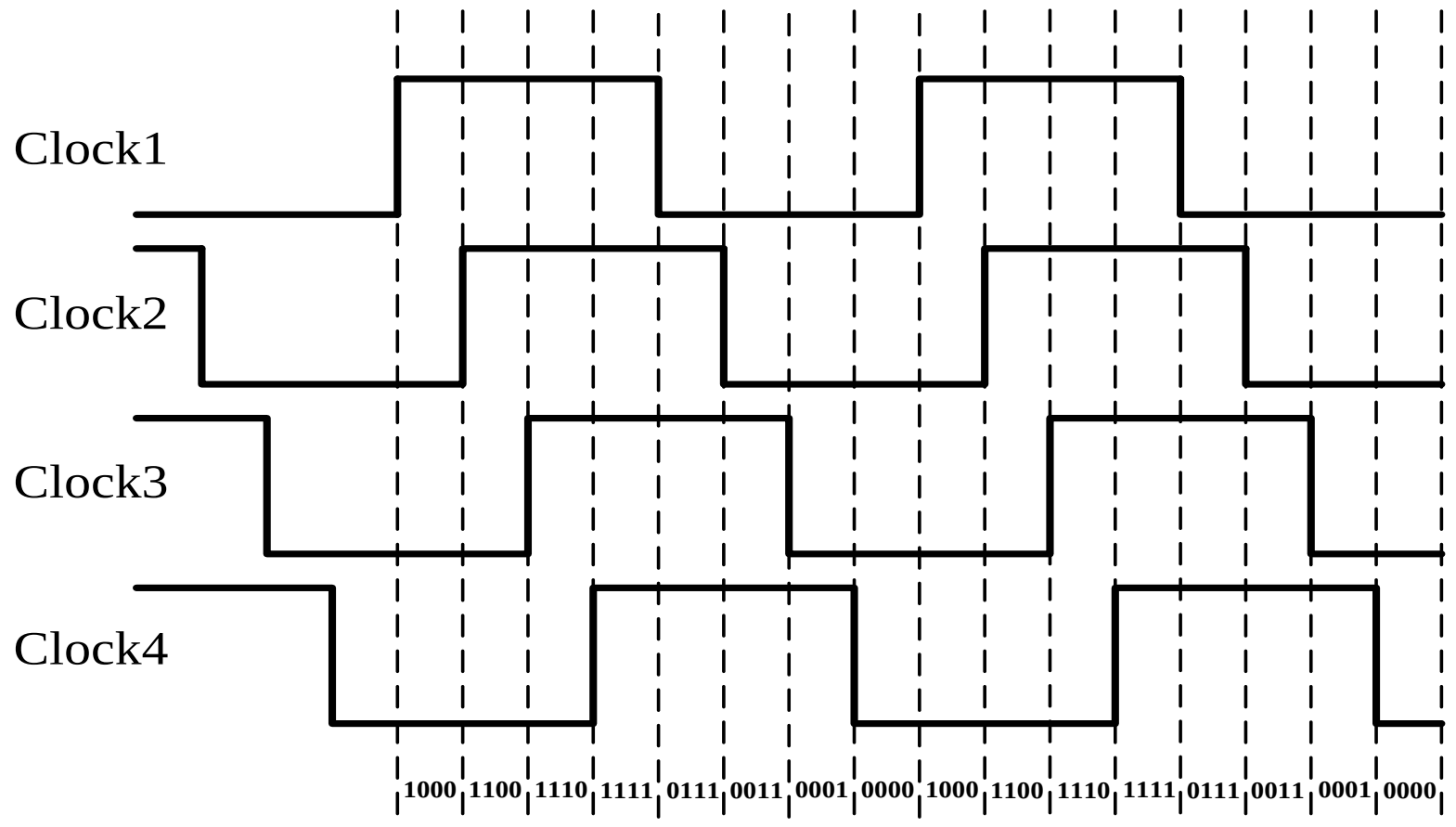

Fig.2 Principle of operation of the interpolation clock

## 3 Architecture

The device implemented for the design was a Stratix III development board, which provides a hardware platform for developing and prototyping low-power, high-performance, and logic-intensive designs. The FPGA series device is EP3CSL150F1152C2 in the development board, and it has eight PLL resources, which can be used for phase shifting and frequency multiplication. There are also two clock oscillators on the board that operate at 125 MHz and 50 MHz. Fig.3 shows a circuit block diagram of the TDC consisting of five components: the PLL, discriminator circuit, coarse module, fine module, and data combination component.

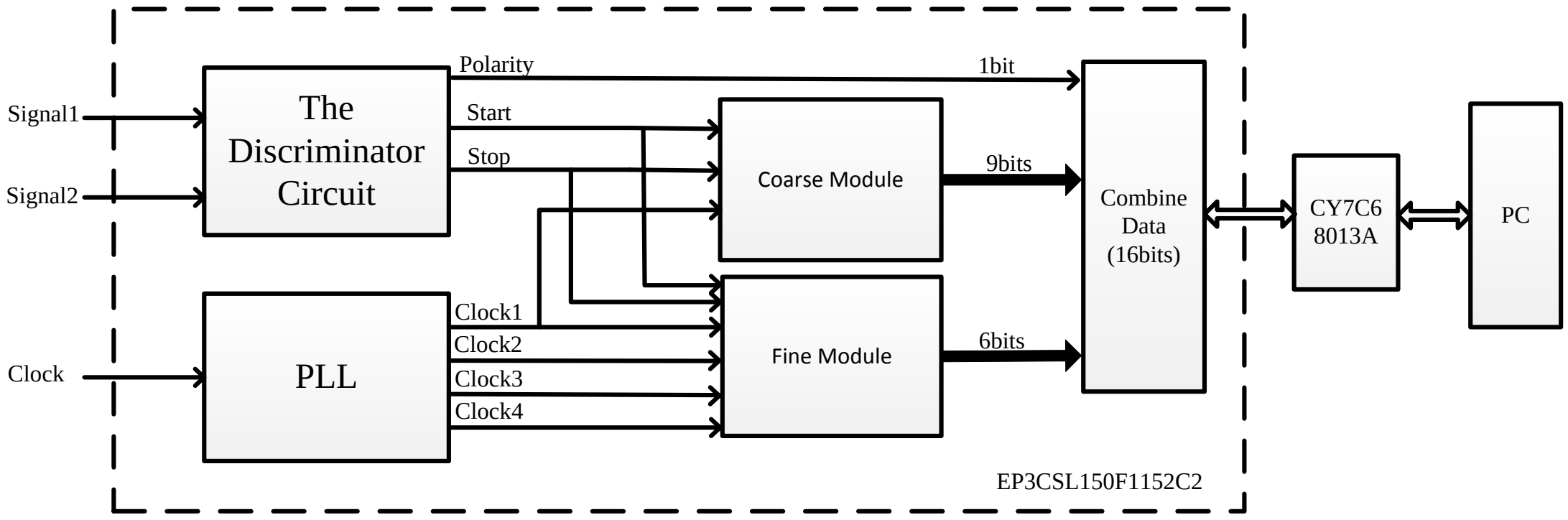

Fig.3 Circuit block diagram of the TDC

The PLL consists of a phase frequency detector, voltage-controlled oscillator, charge pump, and low-pass filter. The PLL has negative feedback-loop characteristics; therefore, the frequency and phase of the output clock generated from the PLL are very stable. Users do not need to design the FPGA-integrated PLL but only need to customize it. A clock oscillator with a frequency of 125 MHz was used as the input clock of the PLL. The clock multiplication factor of the PLL was set to eight; the clock duty cycle was 50%; and the clock phase shifts were set to 0° (clock1), 45° (clock2), 90° (clock3), and 135° (clock4). Thus, the frequency of the four output clocks was 1 GHz, and the phase between adjacent clocks was 45°. Clock1 is also the counting clock of the coarse module. The four clocks make up the interpolation clock, and it divides Clock1 into eight parts, as shown in Fig.2.

### 3.1 Discriminator Circuit

In previous designs of TDCs, the start and stop signals were known. However, we may not be able to determine the start or stop signal in some practical applications. In our design, we add a discriminator circuit to deal with this situation. The circuit comprises a pulse stretching circuit and an identification circuit. The pulse stretching circuit can broaden or shorten the width of a signal pulse, and the identification circuit is used to identity which signal is the first one. With knowledge of only the rising edge of the signal, the circuit can identify the two signals and provide a bit to indicate the polarity. For example, the input signals are signal1 and signal2. If signal1 arrives at the front, the discriminator will output a high level; otherwise, the discriminator will output a low level. A timing simulation diagram of the discriminator circuit is shown in Fig.4.

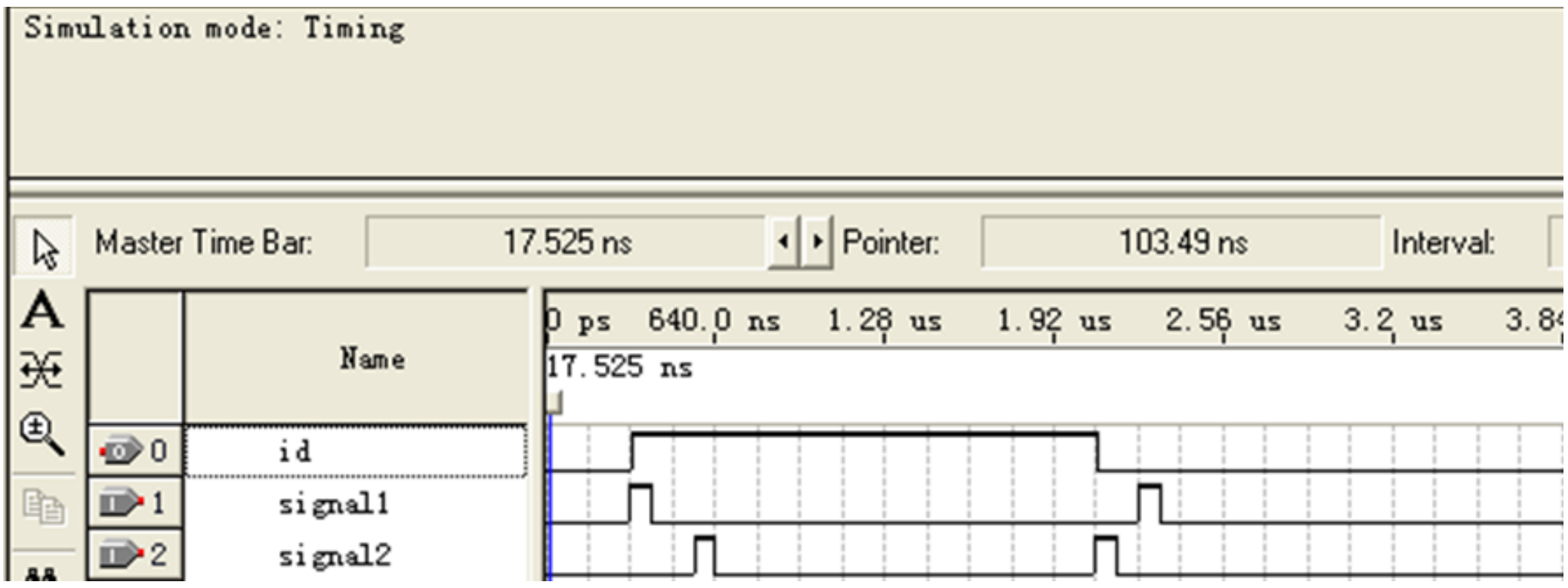

Fig.4 Timing simulation diagram of the discriminator circuit

### 3.2 Coarse Module

The coarse module consists of a Gray-code counter, D flip-flop, and Gray-code-to-binary component. The Gray-code counter has a 9-bit data width and is triggered by Clock1. When a start-signal transition occurs, the counter is sampled by the start register, and the same operation also occurs when the stop signal is delivered to the counter. The time difference between the stop and start registers is the coarse measurement. The coarse module samples the rising edge of the clock after the start and stop signals and detects their phase difference. The stop signal also triggers the D flip-flop to latch the count of the counter. In the Gray-code-to-binary component, counter data are converted into binary form.

### 3.3 Fine Time Module

The fine time module has four D flip-flops and an encoder. Fig.5 shows the architecture of the fine time module. The four D flip-flops are driven by a signal, and their inputs are serially connected with Clock1, Clock2, Clock3, and Clock4. When the rising edge of the signal is present, the D flip-flop can sample the stage of the clock: a logic zero or logic one. The 4-bit data of the D flip-flops are converted into 3-bit data through the encoder. Table1 summarizes the principle of the encoder. In this manner, the range of the coarse counter can be improved.

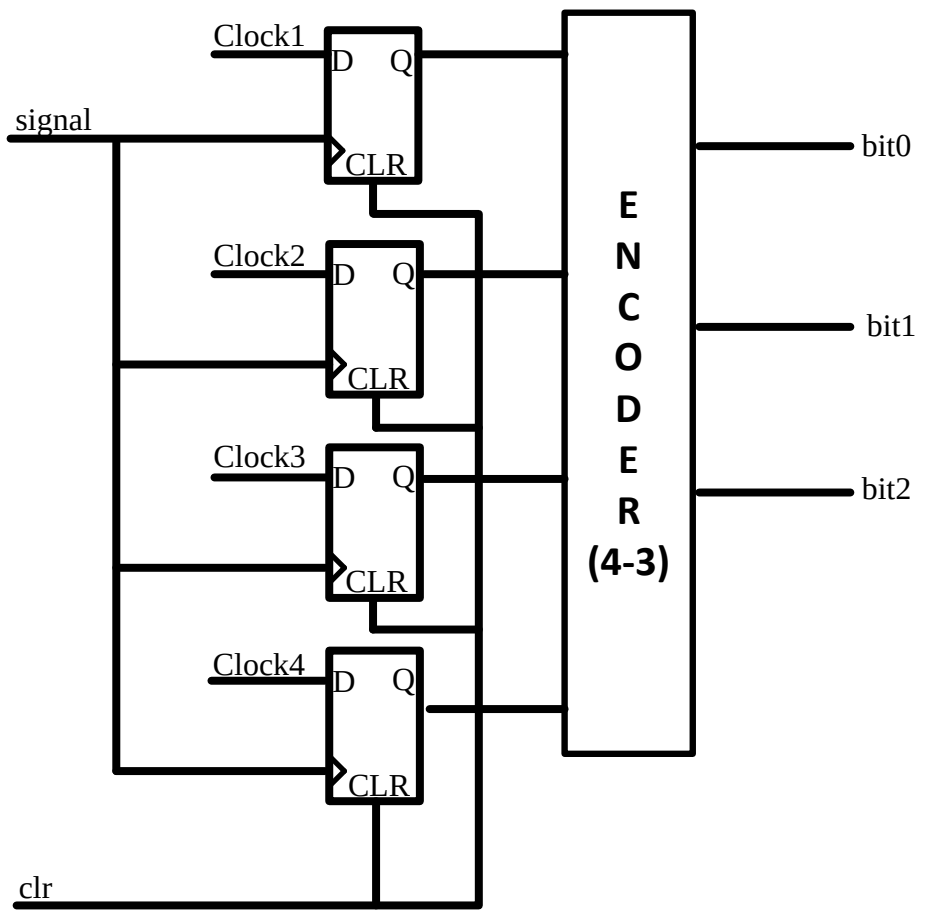

Fig.5 Architecture of the fine time module

Table1. Encoding principle of the encoder.

| output of the D flip-flops | 1000 | 1100 | 1110 | 1111 | 0111 | 0011 | 0001 | 0000 |
|---|---|---|---|---|---|---|---|---|
| Encoder output | 000 | 001 | 010 | 011 | 100 | 101 | 110 | 111 |

### 3.4 Combine Data Component

All of the data will make up 16 bits of data. The most significant digit is the identification data words, and the lower six bits are the fine timing of the start and stop signals. The other bits of data words are the time of coarse module. The 16-bit data words will be sent to a personal computer by USB chip CY7C68013A. The CY7C68013A works in slave first-in–first-out (FIFO) mode in which a buffer is used. The EP2 is configured to be the data in the endpoint of a triple-buffer mode, which stores the data sent to the personal computer from the FPGA[14]. After that, the TDC will be reset and wait for the next time interval.

## 4 Results and discussion

In order to verify our design based on a multiphase clock, a prototype was designed using an FPGA chip. The frequency of the external clock is 125 MHz. Two digital pulse signals were generated from a Tektronix AFG3102 signal generator. The frequency of the generator was set to 1 kHz. Because the TDC does not need to know which signal is a start or stop signal, two signals were assumed to be signal1 and signal2 and connected to the two outputs of the signal generator. The delay of signal1 was set to zero. The time difference between the two pulses can be obtained by setting the delay of signal2 to obtain a different time interval. Fig.7 shows the distribution of the time interval between the two signals, and the standard deviations $\sigma$ of the two distributions are 113 ps and 107 ps in the range of 125 ps. The measurement results are consistent with our theoretical values.

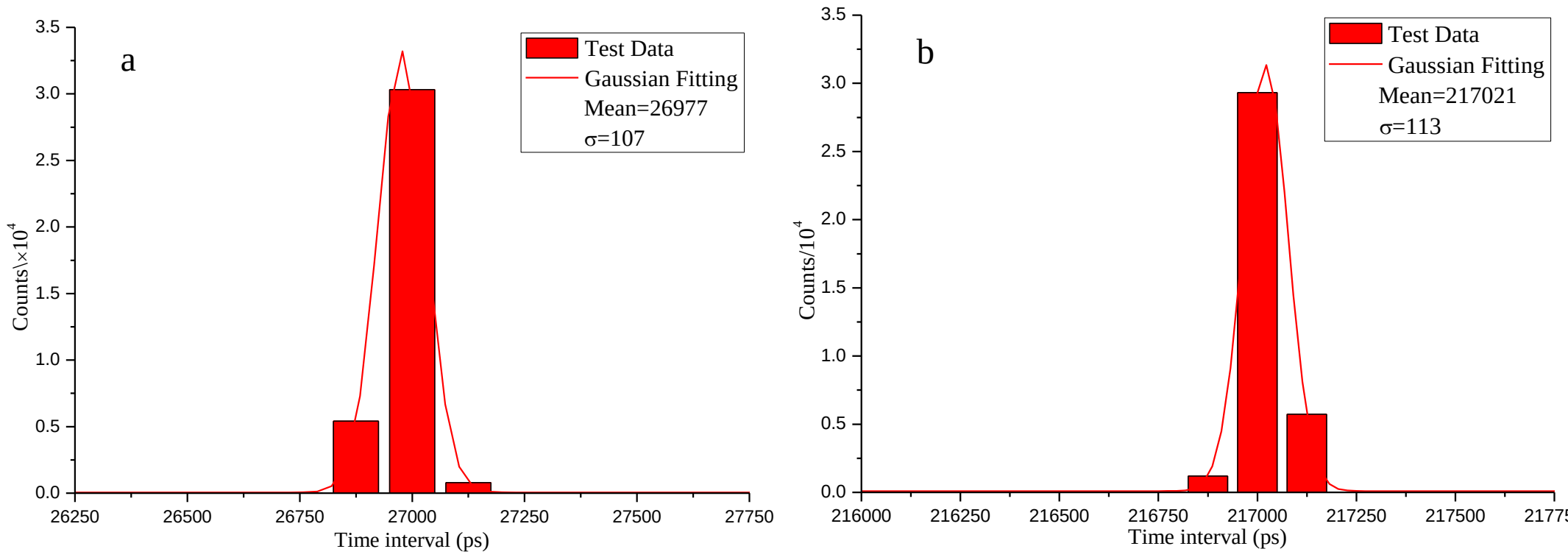

Fig.6 Distributions of the time interval between two signals

The linearity of the TDC was tested by changing the time interval between signal1 and signal2. We measured 30 points in the time interval from 10 ns to 300 ns in increments of 10 ns. The mean value of the measured value could be obtained by performing a Gaussian fit of the distribution of time interval. Fig.7 shows the linear relationship between the measured values and the set values. According to the linear fitting, the correlation coefficient is equal to 1, and the nonlinear error is 0.056%; thus, the linearity of the TDC is very good.

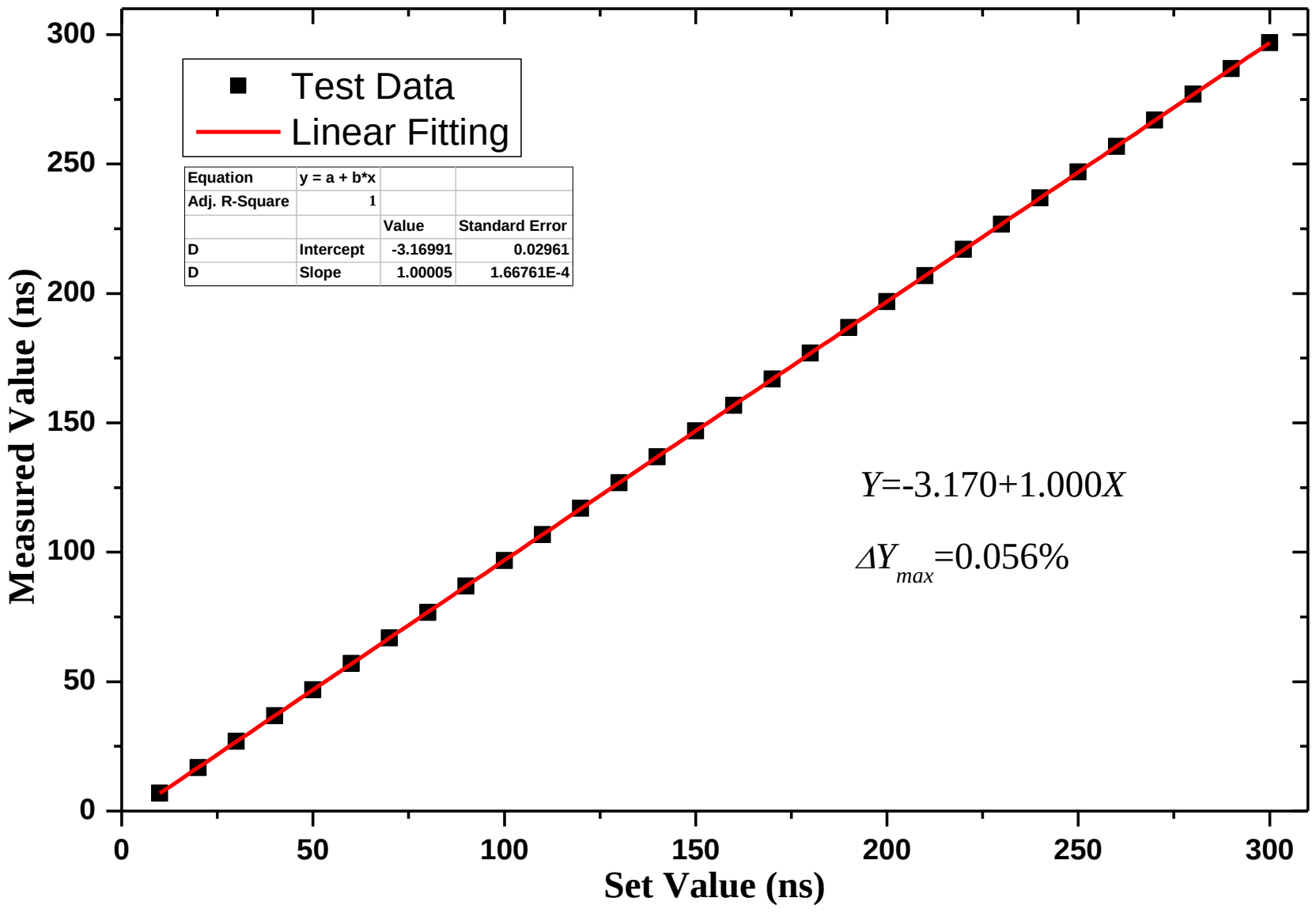

Fig.7 Linearity of the system

## 5 Conclusion

A high-precision TDC was designed using a single FPGA to verify the proposed method of time interpolation by using a multiphase clock generated from a PLL. This time interpolation method can improve the measurement accuracy and reduce the use of FPGA logic resources. The TDC can be easily implemented using an FPGA chip. The resolution of the TDC implemented using an EP3SL1501152C2 device is 125 ps, and the correlation coefficient is equal to 1. Because of the performance of the TDC, it can be used in nuclear experiments and many other applications that need to test for the time interval.